\documentclass[intlimits,twoside,a4paper]{article}

\usepackage[cp1251]{inputenc}
\usepackage[eqsecnum]{cmpj3}

\usepackage{bm}

\issue{2020}{23}{3}{33601}
\doinumber{10.5488/CMP.23.33601}
\title[High pressure study of KBeO3]%
{High-pressure induced magnetic phase transition in half-metallic $\textbf{KBeO}_\textbf{3}$ perovskite %
}
\author[M. Hamlat {\it et al.}]{M. Hamlat\refaddr{label1}, K. Amara\refaddr{label1}, K. Boudia\refaddr{label2},
         F. Khelfaoui\refaddr{label1}\thanks{Corresponding author. E-mail: fraiha\_phys@yahoo.fr}, H. Boutaleb\refaddr{label1} }
\addresses{
\addr{label1} Laboratory of Physico-chemical studies, $\textrm{D}^\textrm{r}$ Tahar Moulay University of Saida, BP 138 Ennasr, 20000 Saida, Algeria
\addr{label2} Laboratory of Physico-chemistry of Advanced Materials, University of Djillali Liabes, 22000 Sidi-Bel-Abbes, Algeria
}
\date{Received October  13, 2019, in final form March 10, 2020}
\begin{document}
\maketitle

\begin{abstract}
In this paper, we present the study of the structural, mechanical, magneto-electronic and thermodynamic properties of the perovskite KBeO$_3$. The calculations were performed by the full potential augmented plane wave method, implemented in the WIEN2k code which is based on density functional theory, using generalized gradient approximation. The computed formation energy and elastic constants indicate the synthesizability and mechanical stability of KBeO$_3$. Moreover, our results showed that the latter is a half-metallic material with half-metallic gap of 0.67 eV and an integer magnetic moment of 3$\mu_{\text{B}}$ per unit cell. In addition, KBeO$_3$ maintains the half-metallic character under the pressure up to about 97 GPa corresponding to the predicted magnetic-phase transition pressure from ferromagnetic to non-magnetic state. The volume ratio $V/V_{0}$, bulk modulus, heat capacity, thermal expansion and the Debye temperature are analyzed using the quasi-harmonic Debye model.
\end{abstract}

\section{Introduction}

The perovskite oxides  system $\textrm{ABO}_\textrm{3}$  has attracted much attention because of their properties, according to the choice of elements A and B, including antiferromagnetism ($\textrm{LaTiO}_\textrm{3}$)~\cite{A1}, ferroelectricity ($\textrm{BaTiO}_\textrm{3}$)~\cite{A2},  ferroelasticity ($\textrm{SrTiO}_\textrm{3}$)~\cite{A3}, antiferroelectricity ($\textrm{PbZrO}_\textrm{3}$)~\cite{A4} , ferromagnetism ($\textrm{YTiO}_\textrm{3}$)~\cite{A5}, half-metal ferromagnets ($\textrm{LiBeO}_\textrm{3}$ \cite{A6}, $\textrm{KMgO}_\textrm{3}$ \cite{A7}). Recently, ferromagnetic materials have been gaining importance in spintronics. Specifically, the focus of the recent research is on the perovskite oxides. Indeed,  several recent works were reported such as  that of Bouadjemi et al. \cite{A8}, where they  found that the cubic perovskite oxide $\textrm{PrMnO}_\textrm{3}$ presents a half-metallic behavior, Rahman and Sarwar \cite{A9} predicted strain- and correlation-induced half-metallic ferromagnetism in BFO ($\textrm{BaFeO}_\textrm{3}$), using DFT calculations. Ali et al. \cite{A10} demonstrated that $\textrm{CaFeO}_\textrm{3}$ is half metal.  However, with its compression up to a certain critical lattice constant, an abrupt change in the electronic and magnetic properties occurs and the compound loses its integer magnetic moment and becomes metallic.
In a recent paper, Khandy et al. \cite{A11} showed that $\textrm{BaNpO}_\textrm{3}$ is half-metallic ferromagnet, using modified Becke and Johnson Gradient (spin) Generalized (mBJ-GGA) and Hubbard potential approximations (GGA+U), and it may be utilized for spintronic devices.
Few researches were carried out on the half-metallic behavior in perovskite oxides without transition (TM) or rare-earth (RE) elements, such as our previous works on $\textrm{LiBeO}_\textrm{3}$ \cite{A6} and $\textrm{KMgO}_\textrm{3}$~\cite{A7}. In focusing on finding out more half-metallic perovskites, without d and f elements, we report a theoretical investigation of the structural, elastic, electro-magnetic and thermodynamic properties of the hypothetical compound $\text{KBeO}_{3}$. We compared our results with those available on the Materials Project database~\cite{A12}. The paper is divided into four sections: after the presented introduction, the second section examines the computational details. The results are reported and discussed in section~3. Finally, the fourth summarizes the results of this work and draws conclusions.

\section{Computational methodology}

In this study, we calculate the structural, elastic, and magneto-electronic properties of perovskite oxide $\text{KBeO}_{3}$ , using the Full Potential Linear Augmented Plane Wave (FP-LAPW), implemented in the code WIEN2k \cite{A13} and within the density functional theory (DFT) \cite{A14}. Semi-relativistic calculations have been made (the spin-orbit effect is neglected). To treat the potential for exchange and correlations, DFT calculations with (PBE-GGA)\cite{A15,A16,A17} are used to determine and obtain the magneto-electronic properties of this compound. To control the convergence of the basis set, the plane wave cut-off value of $R_{\text{MT}}K_{\text{max}} = 8$ is used, and  we chose  muffin-tin radii (MT) of 2, 1.42, 1.33 a.u.  (atomic unit) for K, Be, O respectively. The wave functions inside muffin-tin spheres (MTS) are expanded in terms of spherical harmonics times radial eigenfunctions up to $l_{\text{max}}=10$. In the out of MTS (i.e. interstitial region), the wave functions are expanded in plane waves. For the Fourier charge density expansion of the potential in the interstitial region, $G_{\text{max}}$ of 14 Ry$^{1/2}$ is used. The charge convergence of $10^{-5}e$ is applied as a convergence criterion for all calculations performed.  The density of states DOS is obtained from the calculated Kohn and Sham  eigenvalues on a fine $k$-grid of $14\times14\times14$ in the irreductible Brillouin zone, using the tetrahedron method \cite{A18}.

 To investigate the mechanical stability of $\text{KBeO}_{3}$ in its cubic structure, the IRelast software \cite{A19}, is used to compute the elastic constants. Evidently, for accurate calculations of the elastic properties, a much denser $k$-point grid is necessary. For that reason, we used a $k$-point sampling grid of $22\times22\times22$ in the irreductible Brillouin zone.

\section{Results and discussion}

This part describes and discuses the structural, mechanical, electronic, magnetic and thermodynamic properties of $\text{KBeO}_{3}$. In addition, we study the pressure effect on its HM stability.

\subsection{Equilibrium parameters and mechanical stability}

\begin{figure}[!h]
\centerline{\includegraphics[width=0.3\textwidth]{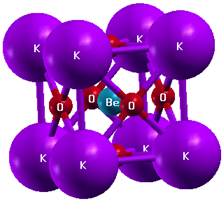}}
\caption{(Color online) Crystal structure of perovskite oxide $\text{KBeO}_{3}$.} \label{fig1}
\end{figure}

As mentioned above, our considered material is the perovskite oxide $\text{KBeO}_{3}$. Therefore, it is investigated in the cubic structure with the Pm3m (221) space group, where the positions of the atoms are K~(0, 0, 0), O (0, 0.5, 0.5) and Be (0.5, 0.5, 0. 5), see figure~\ref{fig1}.

\begin{figure}[!t]
\centerline{\includegraphics[width=0.5\textwidth]{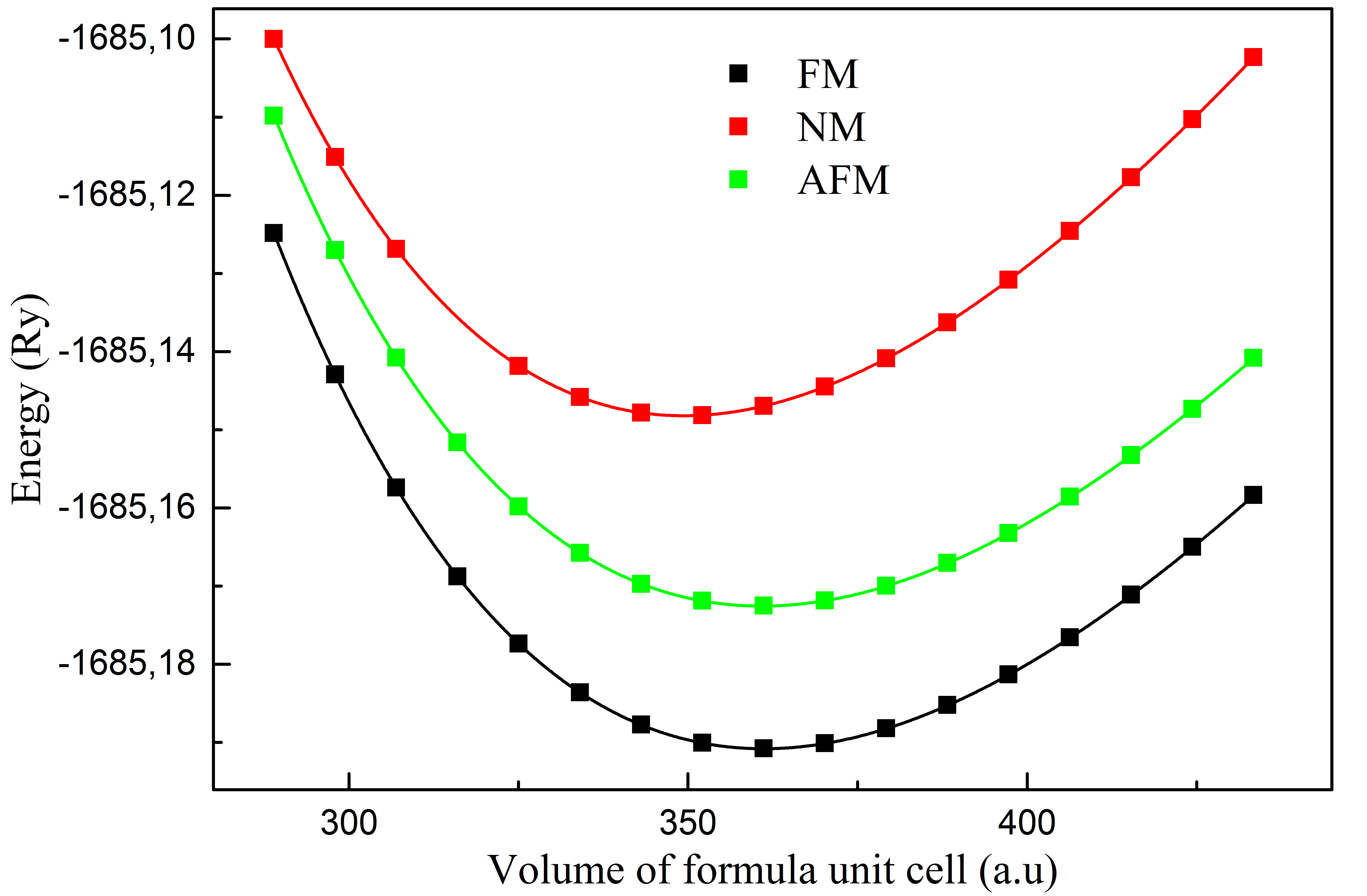}}
\caption{(Color online) Total energy as a function of the volume of the formula unit cell for $\text{KBeO}_{3}$ in its FM, NM and AFM phases.} \label{fig2}
\end{figure}

 In order to determine the magnetic ground state, we calculate the energies for different volumes in the non-magnetic (NM), ferromagnetic (FM), and antiferromagnetic (AFM) phases. The equilibrium lattice parameter, bulk modulus $B$ and its derivative $B'$ are determined by fitting the curve of the total energy as a function of volume to the Birch-Murnaghan equation \cite{A20}, given by the following expression:
\begin{equation}\label{eq1}
    E(V) = E_{0} + \frac{9V_{0}B_{0}}{16} \bigg\{ \bigg[\bigg(\frac{V_{0}}{V}\bigg)^{2/3}-1\bigg]^{3} B_{0} + \bigg[\bigg(\frac{V_{0}}{V}\bigg)^{2/3}-1\bigg]^{2}\bigg[6-4\bigg(\frac{V_{0}}{V}\bigg)^{2/3}\bigg] \bigg\}
\end{equation}
Where $E_{0}$ is the equilibrium total energy, $V_{0}$ is the volume of the unit cell at zero pressure, $B_{0}$ and $B'_{0}$ are the bulk modulus and its pressure derivative, respectively. The $E(V)$ curves are illustrated in figure~\ref{fig2}. From this figure, it can be seen that our compound is more stable energetically in its FM phase. We find the ground-state parameters ($a_{0}\,, B_{0}$ and $B'_{0}$), as shown in table~\ref{table1}. The obtained lattice parameter of FM phase agree very well with that of Persson \cite{A12}.

\begin{table}[htb]
\caption{Calculated values of the equilibrium lattice parameter $a_{0}$ (in {\AA}), bulk modulus $B_{0}$ (GPa) and its pressure derivative, and ground state energy $E_{0}$ (in Ry) for $\text{KBeO}_{3}$.}
\label{table1}
\vspace{2ex}
\begin{center}
\renewcommand{\arraystretch}{0}
\begin{tabular}{|c||c|c|c|c|}
\hline
Phase &	$a_{0}$ &	$B$  &	$B'$ &	$E_{\text{min}}$ \strut\\
\hline
\rule{0pt}{2pt}&&&&\\
\hline
FM	 & 3.7269    &	97.3655	 &  4.2529	&$-1685.148191$ \strut\\
 &3.7716~\cite {A12}&&& \strut\\
\hline
NM	 & 3.7674  & 	87.8309	 &  4.1719	& $-3370.345067$\strut\\
\hline
AFM &	3.7683 &	90.8969	 &  4.2754	& $-1685.190774$\strut\\
\hline
\end{tabular}
\renewcommand{\arraystretch}{1}
\end{center}
\end{table}

\par The mechanical stability of a cubic system requires that Born's stability criteria should be met \cite{A21}: $C_{44}>0$, $C_{11}-|C_{12}|>0$, $C_{11}+2C_{12}>0$  and  $C_{12}<B<C_{11}$, which reflects the stability of our material in this structure against elastic deformations. The calculations of elastic constants  are performed, using the energy-strain approach implemented in the WIEN2k code \cite{A22}, for the energetically favorable phase. The obtained values of the elastic constants $C_{11}, C_{12}$ and $C_{44}$ of the $\text{KBeO}_{3}$ compound, as well as those of \cite{A12}, are listed in table~\ref{table2}.
\par According to Born's stability criteria, our compound is found to be mechanically stable in this phase while its instability is shown by Projector Augmented Wave (PAW) method, as implemented in the Vienna Ab Initio Simulation Package (VASP) software~\cite{A12vasp}.This discrepancy is not due only to different methods used  but can also be explained by the fact that, in the latter work, the used $k$-mesh of $6\times6\times6$ probably leads to inaccurate results and even to the prediction of a mechanically unstable phase.
\par Other quantities related to the elastic constants can be deduced, such as the shear modulus $G$, the Young's modulus $E$, the anisotropic parameter $A$, the ratio $B/G $ and the Poisson's ratio $\nu$. The obtained $B/G$ ratio value  of our material is higher than the critical value 1.75 which separates the ductile/brittle (brittle<1.75<ductile) behavior \cite{A23}, as mentioned in table~\ref{table2}.  Accordingly, our compound can be classified as a ductile material. As can be seen from table~\ref{table2}, the anisotropic parameter $A$ is higher than 1. Thus, $\text{KBeO}_{3}$ exhibits an anisotropic character.
\begin{table}[htb]
\caption{Elastic constants $C_{11},C_{12},C_{44}$, bulk $B$, shear $G$, Young $E$ moduli (in GPa), anisotropic parameter $A$ , $B/G$ ratio,  and Poisson's ratio $\nu$ for $\text{KBeO}_{3}$}
\label{table2}
\vspace{2ex}
\begin{center}
\renewcommand{\arraystretch}{0}
\begin{tabular}{|c||c|c|c|c|c|c|c|c|}
  \hline
 B 	& $C_{11}$ & 	$C_{12}$ & 	$C_{44}$ & 	$G$	 & $E$ & 	$A$ & 	$B/G$ & $\nu$ \strut \\
 \hline
 \rule{0pt}{2pt}&&&&&&&&\\
 \hline
97.180 &	124.0838 &	83.7281	& 45.7706	& 35.533	& 95.018	& 2.2683	& 2.73	& 0.337 \strut \\ \hline
112~\cite{A12}&90~\cite{A12}&124~\cite{A12}&41~\cite{A12}&$-47$~\cite{A12}&&$-5.8$~\cite{A12}&&0.75~\cite{A12}\strut \\ \hline
\end{tabular}
\renewcommand{\arraystretch}{1}
\end{center}
\end{table}
\par To further confirm the stability of this compound, we  calculated the formation energy, in order to examine the thermodynamic stability related to its synthesizability. The formation enthalpy can be computed by the following relation:

\begin{equation}\label{eq2}
                    E_{f}(\text{KBeO}_{3})= E_{\text{tot}}(\text{KBeO}_{3})- E_{\text{Bulk}}(\text{K})- E_{\text{Bulk}}(\text{Be})-3/2 E(\text{O}_{2})
\end{equation}
where  $E_{\text{Bulk}}(\text{K}), E_{\text{Bulk}}(\text{Be})$, and $E(\text{O}_{2})$, correspond to the total energy for K, and Be, per atom, and for $\textrm{O}_{2}$ molecule, respectively. The obtained negative formation energy of $-0.056$ Ry indicates that  $\textrm{KBeO}_{3}$ is thermodynamically stable. Therefore, it can be synthesized in the normal conditions.

\subsection{Electronic properties}

\begin{figure}[!t]
\centerline{\includegraphics[width=0.5\textwidth]{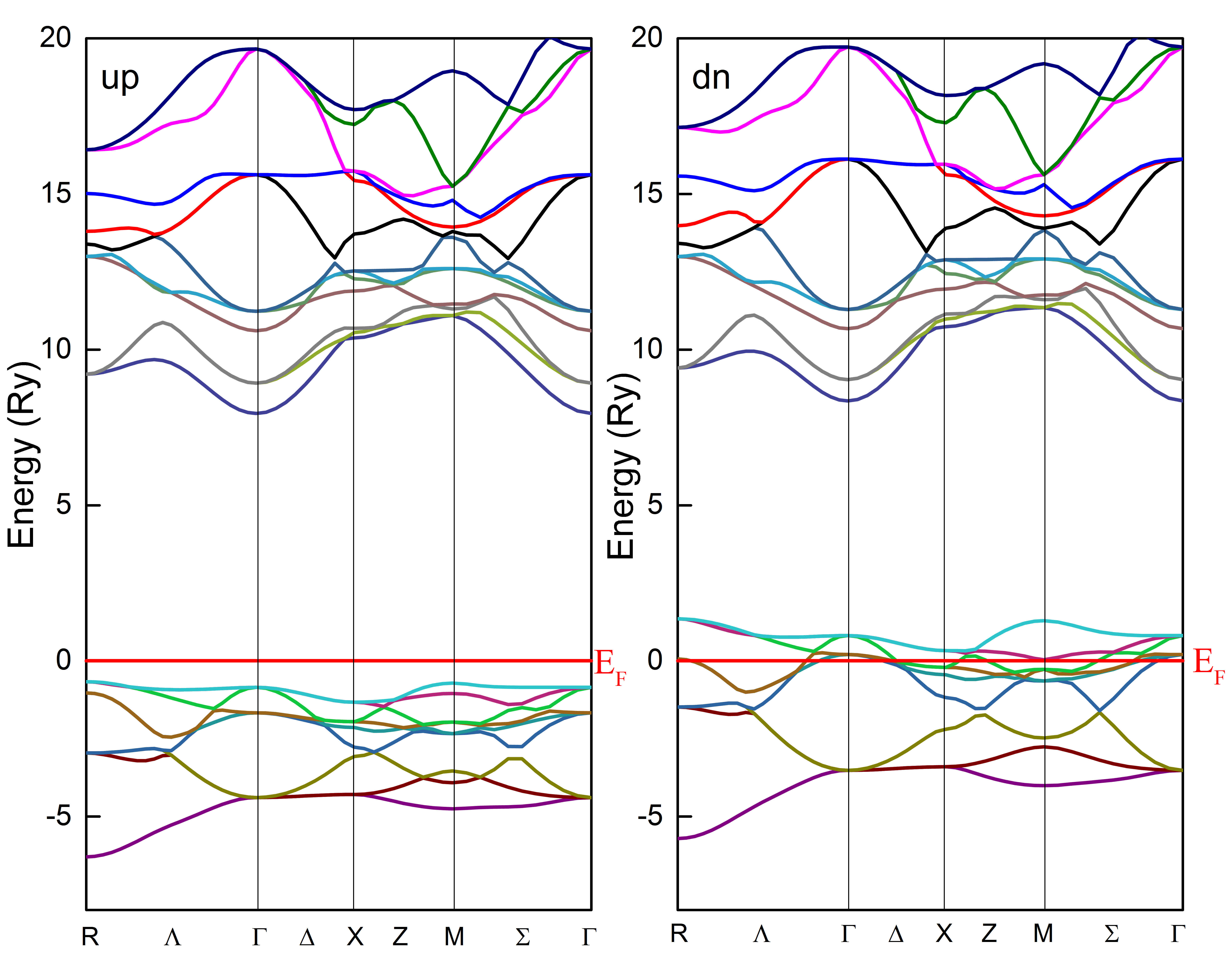}}
\caption{(Color online) Calculated spin-polarized band structures for $\text{KBeO}_{3}$ at its equilibrium lattice constant.} \label{fig3}
\end{figure}

The importance of these properties makes it possible to analyze and understand the nature of the bonds between the different elements of a material. Band structures allow one to understand the phenomenon of half-metallic character in a compound. In this part, we performed electronic structure calculations for $\text{KBeO}_{3}$ where we use the equilibrium lattice constant corresponding to the FM phase. We used GGA, to calculate spin-polarized band structures, as shown in figure~\ref{fig3}. It is clear that in the spin-dn band structure, the valence  band (VB) and Fermi level overlap (i. e., intersect). This indicates that the compound shows a metallic behavior for this spin direction, whereas the spin-up band structure has a gap of 8.61 eV, which represents an insulating character. Therefore, $\text{KBeO}_{3}$ is half-metallic compound.

\par
\begin{figure}[!t]
\centerline{\includegraphics[width=0.5\textwidth]{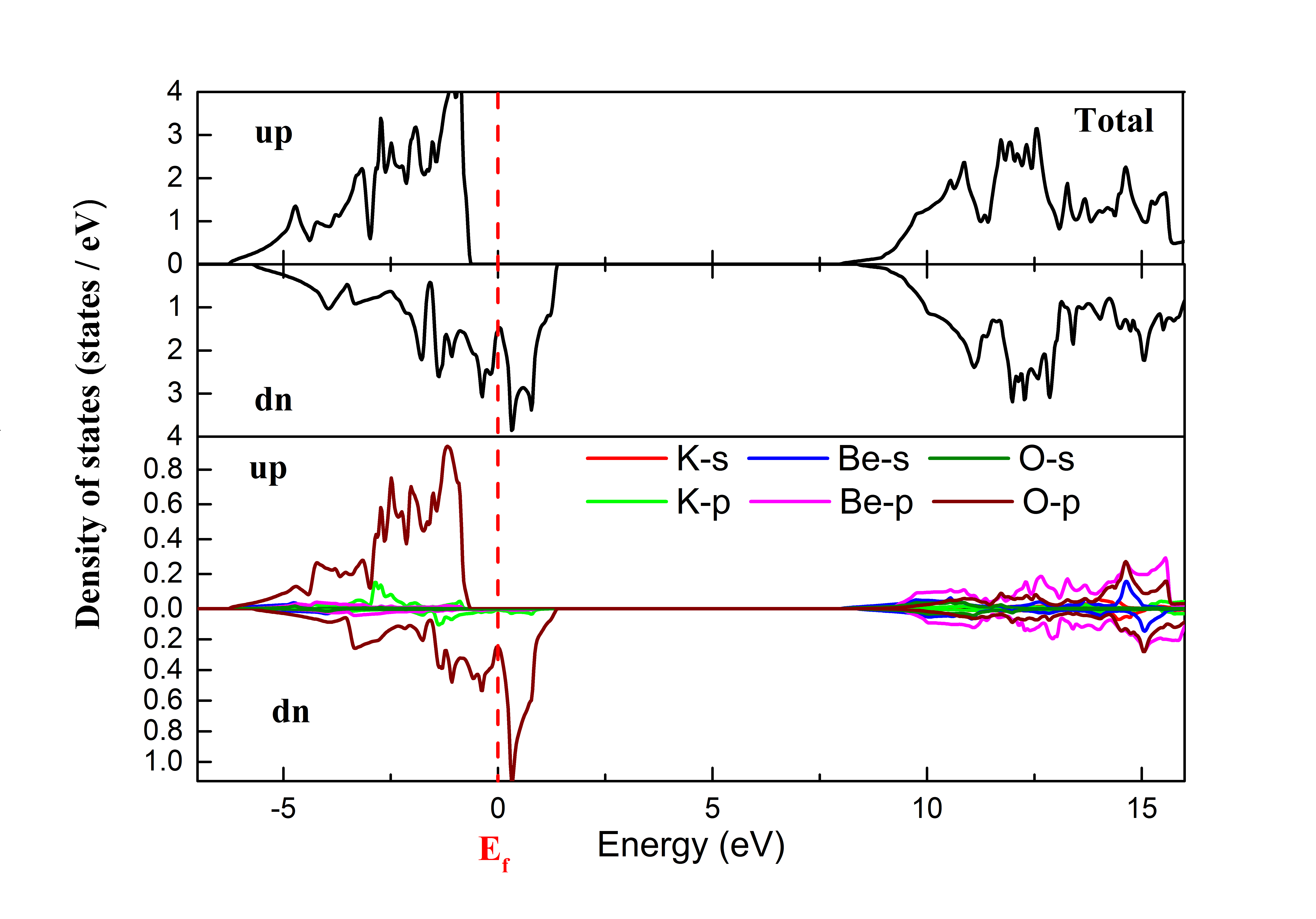}}
\caption{(Color online) Total and partial spin-dependent densities of states for $\text{KBeO}_{3}$ at its equilibrium lattice constant.} \label{fig4}
\end{figure}

 The total and partial densities of state for perovskite oxide $\text{KBeO}_{3}$ in its ferromagnetic phase are calculated at its structural equilibrium state using GGA, as illustrated in figure~\ref{fig4}. From the total and partial densities, found by the FP-LAPW method, we can distinguish three important energy intervals. The first region is located at about [$-7$ eV, $-0.63$ eV] in both spin directions, the contribution of O-$p$ orbital is dominant. In the second interval, there is a band gap of 8.61 eV in the spin-up around Fermi level. The conduction band, between [7.95 eV, 16 eV] is formed mainly by  $p$ orbitals of (K) ,(Be) and (O) atoms.

\subsection{Magnetic properties}

In this section, we begin to understand the magnetic character of $\text{KBeO}_{3}$ compound. As shown in table~\ref{table3}, the magnetic character comes mainly from O atoms. It can be seen that the main contribution to the magnetic character in the resulting DOS, obtained by the GGA method around the Fermi level, is due to the orbital $p$ of O atoms. It is found that O has a larger magnetic moment. This result is confirmed by the results of the DOS (figure~\ref{fig4}). The total magnetic moment of our studied compound has the integer value of 3$\mu_{\text{B}}$. Such as our previously studied compounds $\textrm{LiBeO}_\textrm{3}$ and $\textrm{KMgO}_\textrm{3}$ \cite{A6,A7},  the magnetic moment value of $3.0 µ_{\text{B}}$ can be clarified by the existance of  21 valence electrons in $\text{KBeO}_{3}$. The single valence electron of K and two electrons of Be are attracted by the Oxygen atoms. Therefore, the $\text{KBeO}_{3}$ valence electrons are distributed as follows: two electrons fill $s$-states of the O atoms and five electrons occupy their $p$-orbitals, giving rise to one unpaired electron for each O atom, leading to a total magnetic moment of $3.0 µ_{\text{B}}$.
In contrast to transition and rare-earth-based perovskites  such as $\textrm{BaFeO}_\textrm{3}$, $\textrm{PrMnO}_\textrm{3}$ and $\textrm{BaNpO}_\textrm{3}$ \cite{A8,A9,A11} with total magnetic moments of 6$\mu_{\text{B}}$, 4$\mu_{\text{B}}$, and 3$\mu_{\text{B}}$, respectively, where the magnetism is originated from TM and RE elements, the contribution of the three O atoms of $\text{KBeO}_{3}$ ensures a comparable total magnetic moment.

\begin{figure}[!t]
\centerline{\includegraphics[width=0.4\textwidth]{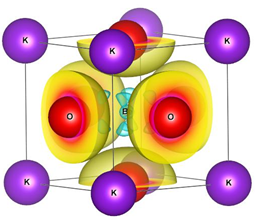}}
\caption{(Color online) Spin density of $\text{KBeO}_{3}$.} \label{fig5}
\end{figure}

 To better highlight the origin of the magnetic moment of our considered compound, the spin density isosurfaces were plotted as shown in figure ~\ref{fig5}. It is clear from this figure that the spin magnetic moment is localized around O atoms which are shown to be ferromagnetically coupled, according to the yellow isosurfaces. Meanwhile, a part of spin magnetic moment around Be atom, as shown by blue drop shapes, indicates  a small negative spin magnetic moment. On the other hand, there is no spin accumulated around K, signifying a negligible magnetic moment value of K.
\begin{table}[!h]
\caption{Calculated total, atom-resolved, and interstitial magnetic moments (in $\mu_{\text{B}}$) in the unit cell for $\text{KBeO}_{3}$}
\label{table3}
\vspace{2ex}
\begin{center}
\renewcommand{\arraystretch}{0}
\begin{tabular}{|c||c|c|c|c|}
  \hline
$M_{\text{tot}}$ &	$M_{\text{K}}$ & 	$M_{\text{Be}}$ &	$M_{\text{O}}$ & 	$M_{\text{int}}$ \strut \\
\hline
\rule{0pt}{2pt}&&&&\\
\hline
3.00068	& 0.01196 &	$-0.01417$ & 	 0.74613 &	0.76451 \strut \\ \hline
\end{tabular}
\renewcommand{\arraystretch}{1}
\end{center}
\end{table}

\subsection{Pressure effect on the half-metallic character}

\begin{figure}[!h]
\centerline{\includegraphics[width=0.6\textwidth]{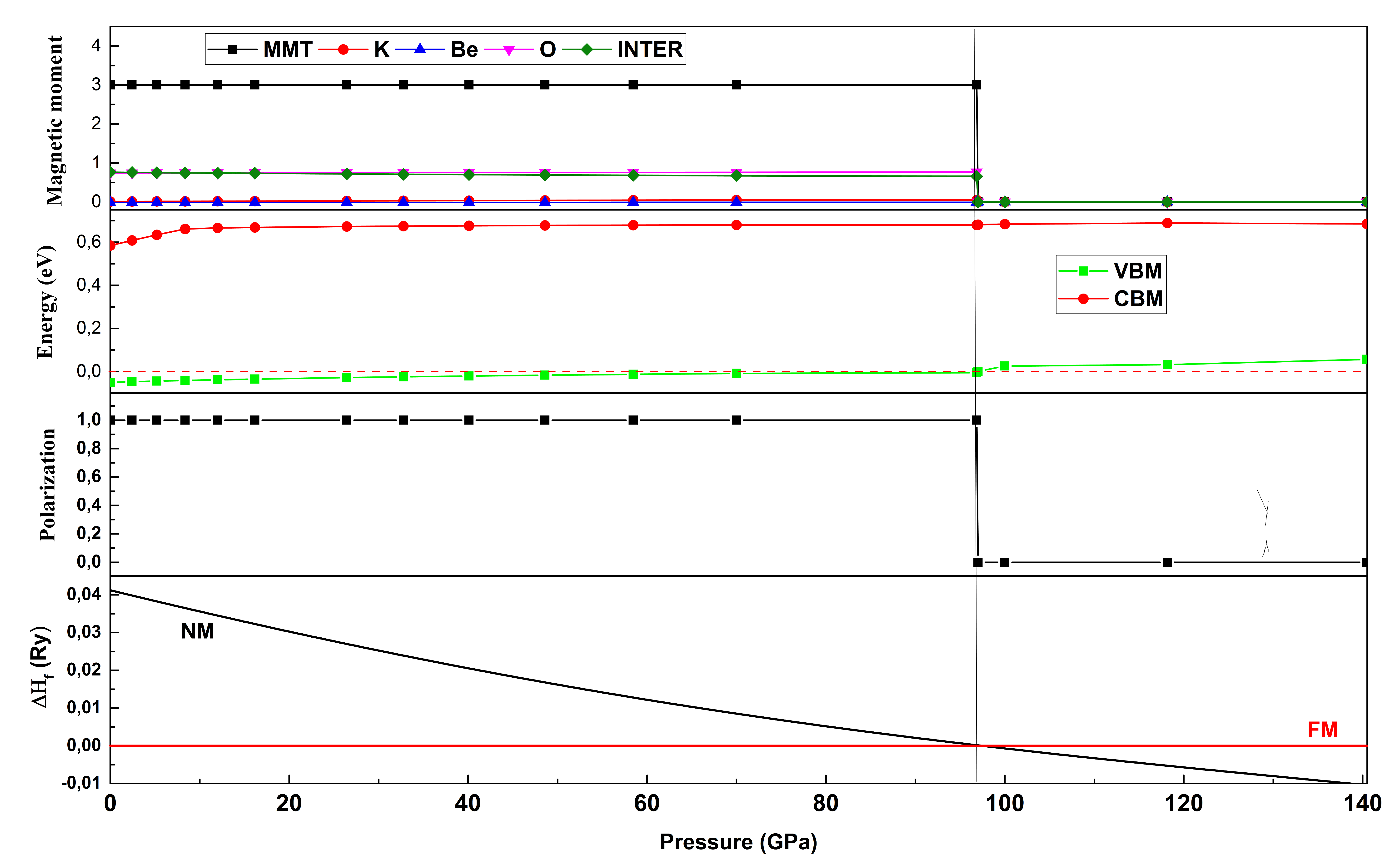}}
\caption{(Color online) Total, atomic, interstitial magnetic moments, CBM, VBM in the spin-up channel, spin polarization, and enthalpy as function of pressure for $\text{KBeO}_{3}$.} \label{fig6}
\end{figure}

Figure~\ref{fig6} shows the total, atomic, interstitial magnetic moments, valance band maximum, conduction band minimum in the spin-up direction, and spin-polarization as a function of the pressure for $\text{KBeO}_{3}$. It can be seen from this figure that the total magnetic moment maintains the value of 3$\mu_{\text{B}}$, once the pressure reached 97 GPa, which corresponds to the lattice parameter value of 6.14 a.u., the total magnetic moment becomes zero 0 $\mu_{\text{B}}$. At this pressure, a magnetic phase transition from FM to NM state may  occur. This result agrees with the  phase transition studied, as presented in figure~\ref{fig6}. At 0 K, the datum of the pressure for the phase transition can be concluded from the ordinary condition of the same enthalpy $H=E+PV$. Regarding $\text{KBeO}_{3}$ compound, the variation of the enthalpy as a function of the pressure at 0~K is noted in figure~\ref{fig6}, the crossing of enthalpy-pressure curves show that the value of transition pressure from FM to NM is about 97 GPa.
The valence band maximum (VBM) and conduction band minimum (CBM) values in the spin-up direction are used to examine the effect of the pressure on the half-metallic stability for our  compound studied (see figure~\ref{fig6}).

The spin polarization (P) at Fermi $E_{F}$ level is related to spin-polarized DOS by the expression:

\begin{equation}\label{eq3}
   P(E_{F}) = \frac{|\rho_{\uparrow}(E_{F})-\rho_{\downarrow}(E_{F})|}{\rho_{\uparrow}(E_{F})+\rho_{\downarrow}(E_{F})}.
\end{equation}
Where $\rho_{\uparrow}(E_{F})$ and $\rho_{\downarrow}(E_{F})$ are the majority and minority state densities at Fermi level $E_{F}$. As is clear from figure \ref{fig6}, the spin polarization at Fermi level is 100\% up to the pressure of  97 GPa. As shown in this figure, it can be seen that the half-metallicity character can be conserved up to 97 GPa. When the pressure value is bigger than 97 GPa, the half-metallicity is lost and the compound becomes a non-magnetic metal.

\subsection{Thermodynamic properties}
The thermodynamic properties of $\text{KBeO}_{3}$ are obtained by means of the quasi-harmonic Debye model~\cite{A24,A25}. The non-equilibrium Gibbs function $G (V, P, T)$ is given by:

\begin{equation}\label{eq3}
   G^{*} (V,P,T)=E(V) + PV + A_{\text{vib}}[\theta(V),T]
\end{equation}
Where $E(V)$ is the total energy of the  unit cell for $\text{KBeO}_{3}$, $PV$ is the constant hydrostatic pressure condition and the $A_{\text{vib}}$  is
 the vibrational Helmholtz free energy, expressed by \cite{A24}:
\begin{equation}\label{eq3}
 A_{\text{vib}} (\theta_{D},T)=n k_{\text{B}} T \left\lbrace 9 \theta_D/8T+3 \ln[1-\text{e}^{(-\theta_{D}/T)}]-D(\theta_{D}/T)\right\rbrace 
\end{equation}
where $n$ is the number of atoms per formula unit and the Debye integral $D(\theta_{D}/T)$, for an isotropic solid, is  \cite{A24}:
 \begin{equation}\label{eq3}
  \theta_{D}= \hbar/k_{\text{B}}(6 \piup^{2} n V ^{1/2})^{1/3} f(\sigma)(B_{s}/M)^{1/2}.
\end{equation}
Where, $M $ is the molecular mass per unit cell.  $B_{s}$ is the adiabatic bulk modulus, which is approximately given by static compressibility \cite{A24}:
 \begin{equation}\label{eq3}
 B_{s}= B(V) = V  \frac{\partial^{2} E(V)}{\partial V^{2}}
\end{equation}
and  $f(\sigma)$ is given by \cite{A24}:
\begin{figure}[!t]
\centerline{\includegraphics[width=0.45\textwidth]{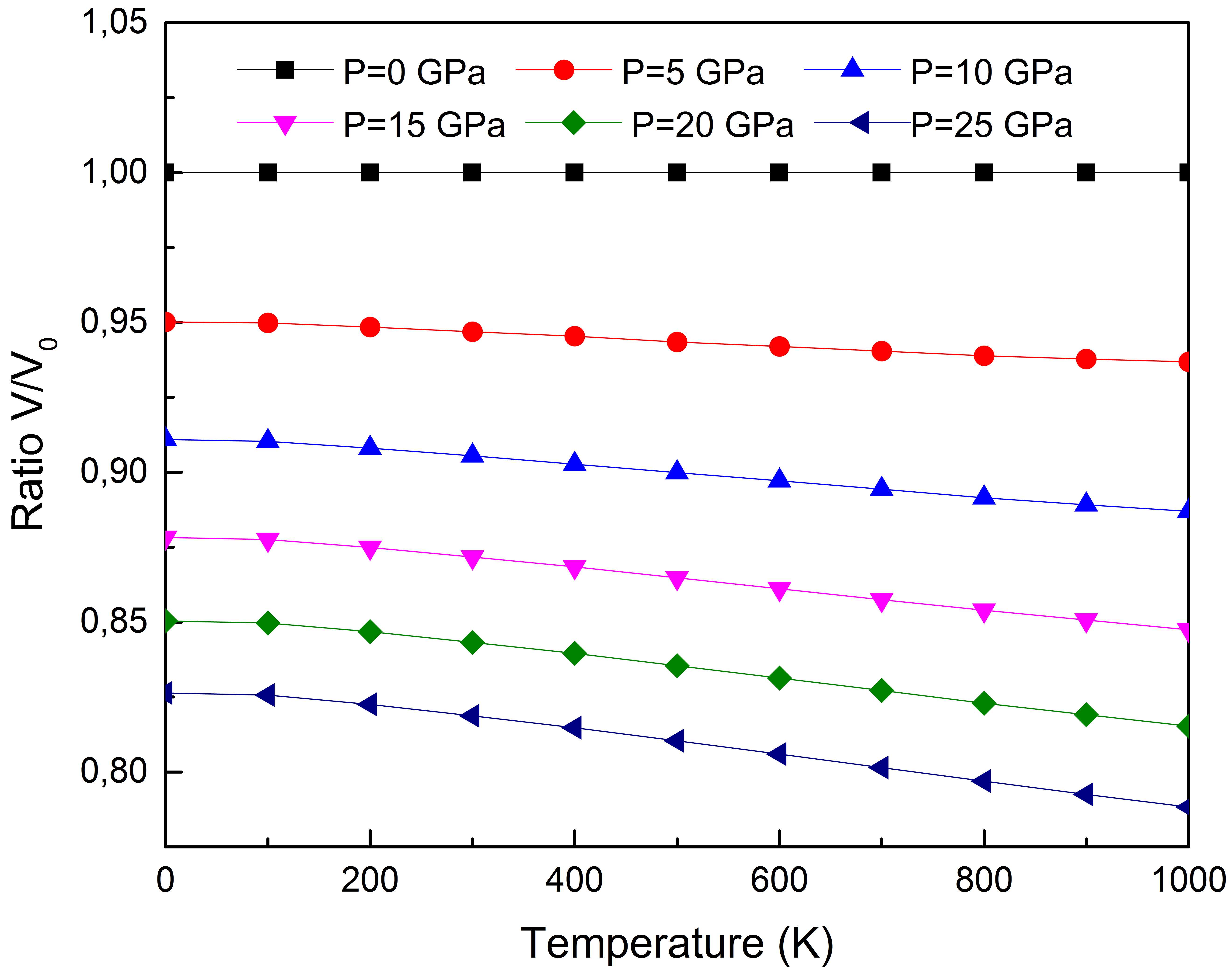}}
\caption{(Colour online) Pressure and temperature dependence of the relative volume $V/V_{0}$ for $\text{KBeO}_{3}$ ($V_{0}$ is the equilibrium volume).} \label{fig7}
\end{figure}
\begin{figure}[!t]
\centerline{\includegraphics[width=0.45\textwidth]{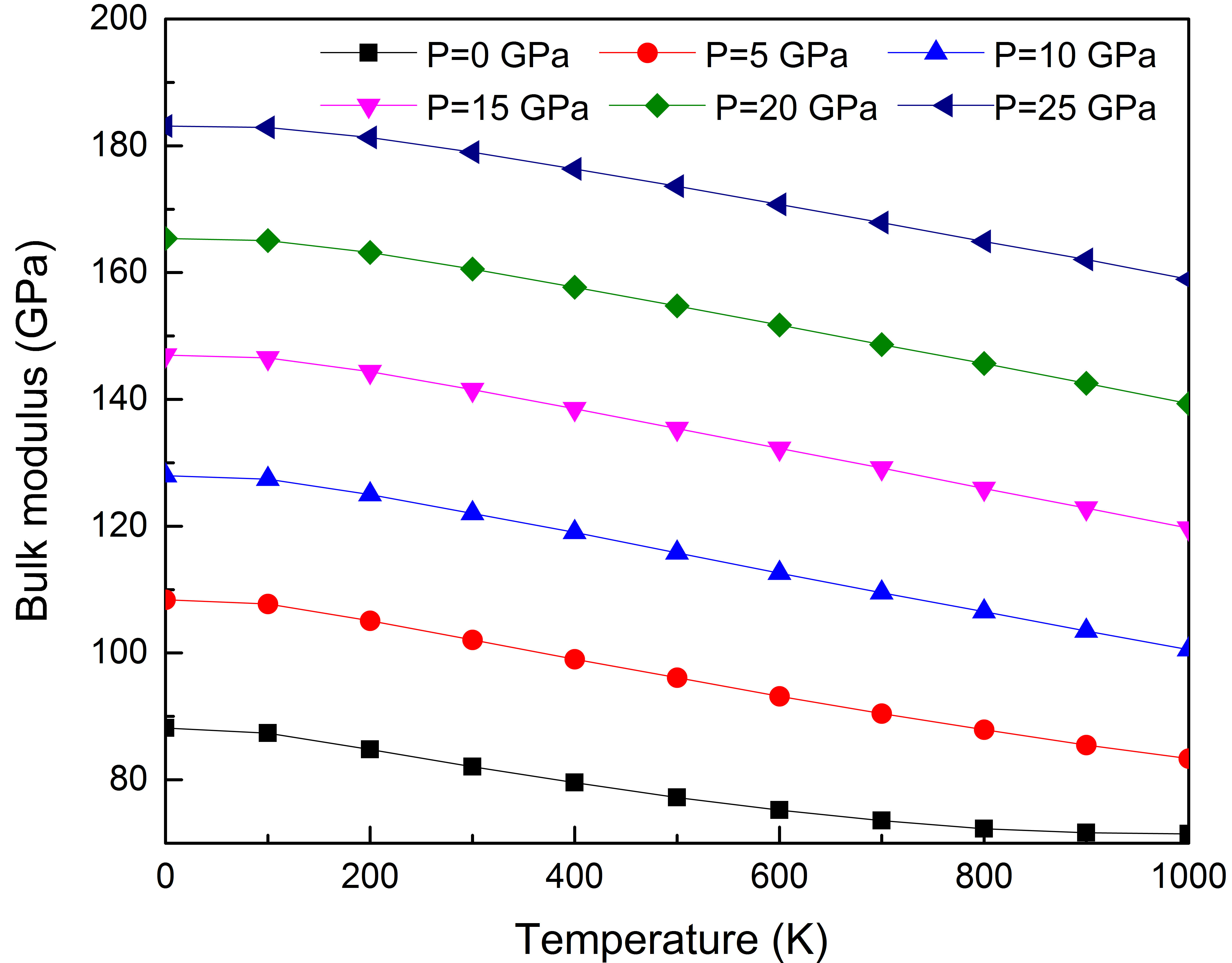}}
\caption{(Colour online) Pressure and temperature dependence of the bulk modulus for $\text{KBeO}_{3}$.} \label{fig8}
\end{figure}
\begin{equation}\label{eq3}
f(\sigma)=\bigg\{3 \bigg[ 2 \bigg(\frac{21+\sigma}{31-2\sigma}\bigg)^{3/2}+ \bigg(\frac{11+\sigma}{31-\sigma}\bigg)^{3/2}\bigg]\bigg\}^{1/3}
\end{equation}
where $\sigma$ is Poisson ratio. The non-equilibrium Gibbs function $G^{*} (V,P,T)$, as a function of $(V, P, T)$ can be minimized with respect to volume $V$ as:
\begin{equation}\label{eq3.9}
\bigg[\frac{\delta^{2} G^{*}(P,V)}{\delta V^{2}}\bigg]_{P,V}=0
\end{equation}
can be obtained by equation~(\ref{eq3.9}). The isothermal bulk modulus $B_{T}$ is given by \cite{A25}:
\begin{equation}\label{eq3}
B_{T}(P,T)=V \bigg[\frac{\delta^{2} G^{*}(P,V,T)}{\delta V^{2}}\bigg]_{P,T}
\end{equation}
The thermodynamic quantities such as heat capacity at a  constant volume $C_{V}$ and entropy $S$ are calculated by the relations\cite{A24,A25}:
\begin{equation}\label{eq3}
C_{V}=3nk_{\text{B}} \bigg[ 4D \bigg(\frac{\theta_{D}}{T}\bigg)- \frac{3\theta_{D}/T}{\text{e}^{\theta_{D}/T}-1}\bigg]
\end{equation}
\begin{equation}\label{eq3}
S_{V}=nk_{\text{B}}\bigg[4D \bigg(\frac{\theta_{D}}{T}\bigg)-3 \ln (1-\text{e}^{-\theta_{D}/T})\bigg]
\end{equation}
where the thermal expansion coefficient $\alpha$ is given by:
\begin{equation}\label{eq3}
\alpha =\frac{\gamma C_{V}}{B_{T}T}
\end{equation}
The thermal properties of $\text{KBeO}_{3}$, in its FM phase, are determined in the temperature range 0--1000~K for different pressures in the range 0--25 GPa. It is evident from figure~\ref{fig7} that the volume ratio $V/V_{0}$ decreases linearly with an increasing temperature at a given pressure and decreases with an increasing pressure at a given temperature.  The bulk modulus relates to the change in the volume of an object, when subjected to pressure changes, and is used to describe hardness \cite{A26}. Figure~\ref{fig8} presents the variation of bulk modulus with temperature for particular pressures (0, 5, 10, 15, 20, 25~GPa). It is clearly seen that $B$ is nearly stable between 0 to 100 K and decreases linearly with an increasing temperature. On the other hand, the compound loses its hardness with an increasing temperature. The bulk modulus, at zero pressure and $T$=300 K, is 82.07~GPa.
\begin{figure}[!t]
\centerline{\includegraphics[width=0.45\textwidth]{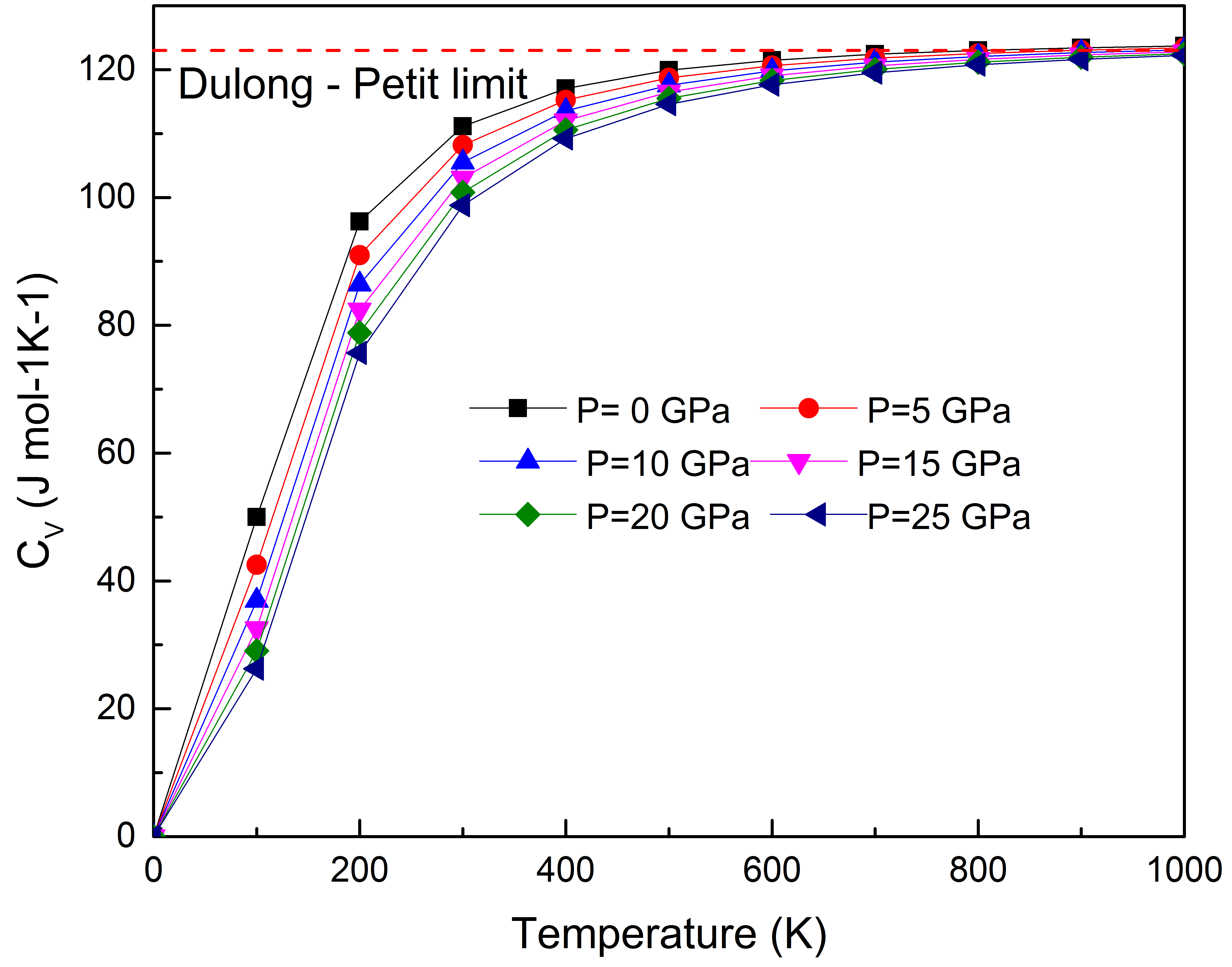}}
\caption{(Colour online) Pressure and temperature dependence of the heat capacity $C_{V}$  for  $\text{KBeO}_{3}$.} \label{fig9}
\end{figure}
\begin{figure}[!t]
\centerline{\includegraphics[width=0.45\textwidth]{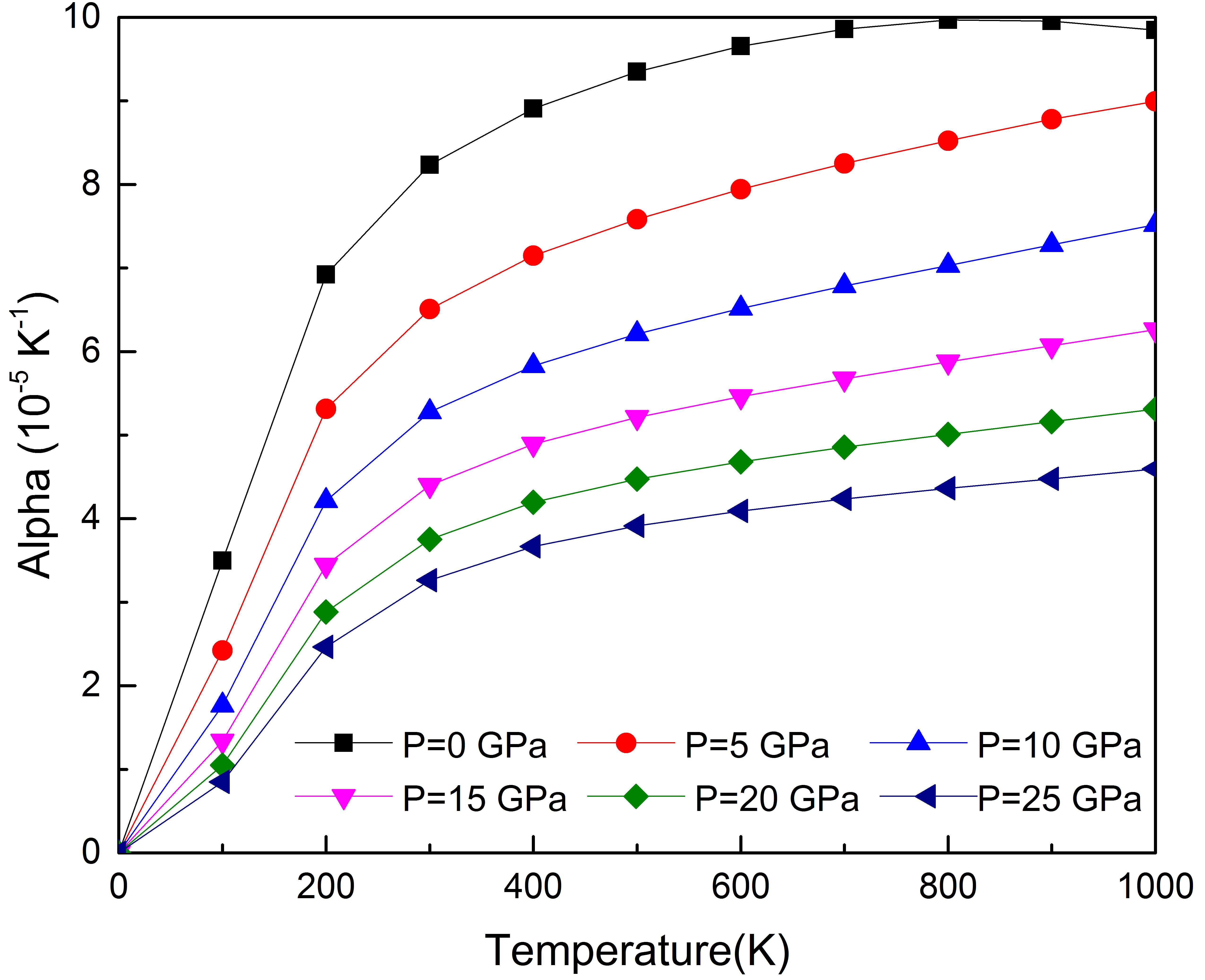}}
\caption{(Colour online) Pressure and temperature dependence of the thermal expansion for $\text{KBeO}_{3}$.} \label{fig10}
\end{figure}
\begin{figure}[!t]
\centerline{\includegraphics[width=0.45\textwidth]{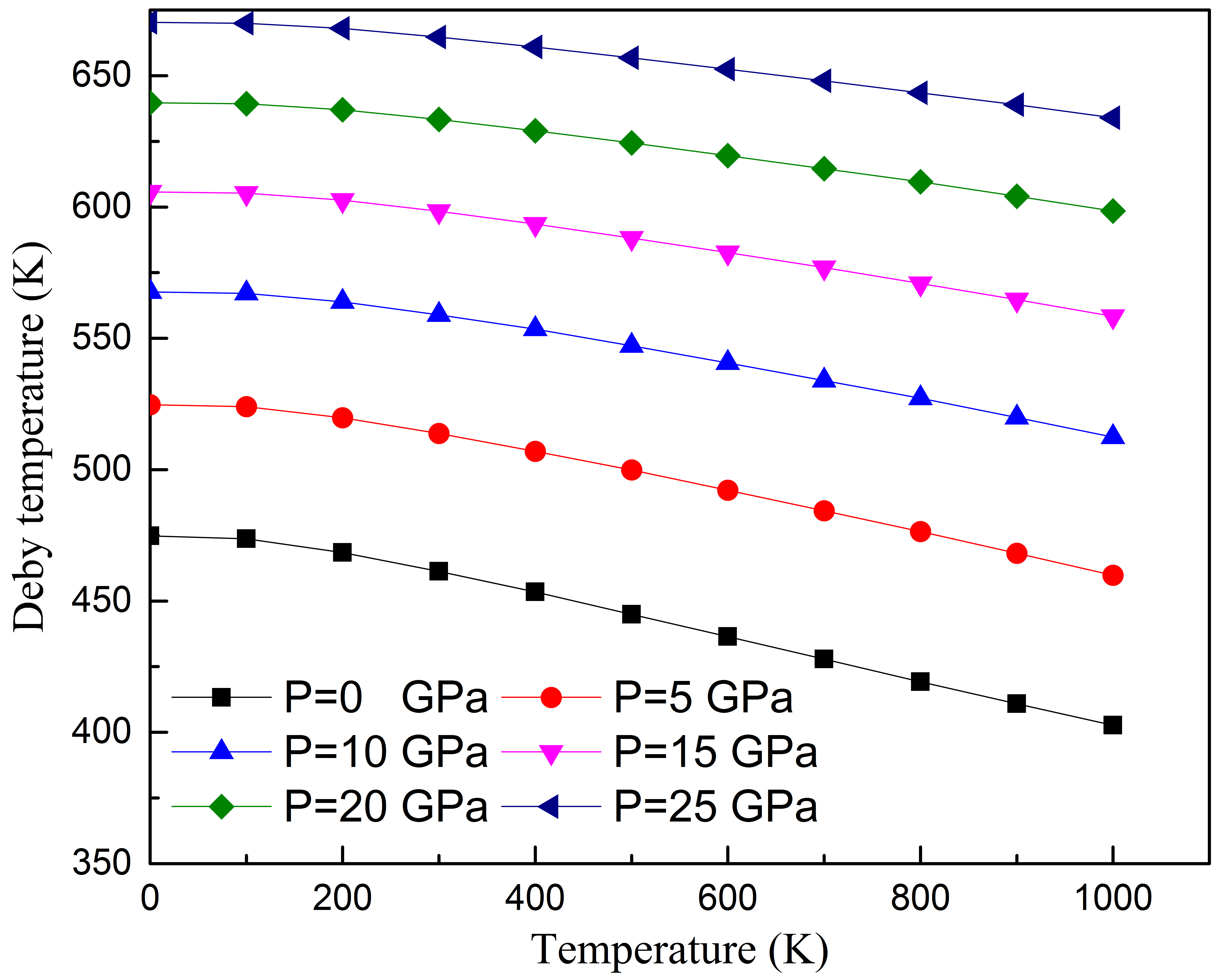}}
\caption{(Colour online) Pressure and temperature dependence of the Debye temperature for $\text{KBeO}_{3}$.} \label{fig11}
\end{figure}

Figure~\ref{fig9} presents an examination for the relation between heat capacity at a constant volume ($C_{V}$) and temperature for 0, 5, 10, 15, 20, and 25 GPa pressures. At a lower temperature, $C_{V}$  values grow rapidly ($T<200$~K) and its evaluation is proportional to $T^{3}$ \cite{A26}, then increases heavily at a higher temperature and approaches the Dulong-Petit limit \cite{A24}. Beyond about 800 K, $C_{V}$ stays the same, its value is about 123~(J/mol~K).

Another thermodynamic parameter for $\text{KBeO}_{3}$, the thermal expansion, is shown in figure~\ref{fig10}, as a function of the temperature and pressure. It can be seen that the thermal expansion coefficient $\alpha$ increases pointedly with an increasing temperature up to 200 K. For higher temperature ($T>200$~K), $\alpha$ progressively approaches a linear increase with temperature, which means that the temperature dependence of alpha is relatively low at high temperatures. At 300 K and zero pressure, the thermal expansion coefficient $\alpha$ value is 8.23$\times10^{-5}$~K$^{-1}$ \cite{A27}.

Figure~\ref{fig11} depicts the temperature and pressure effects on Debye temperature. From this figure, it can be seen that $\theta_{D}$ is nearly stable from 0 to 100 K and decreases with increasing the temperature. Moreover, it is evident that when the temperature is constant, the Debye temperature increases with the pressure applied. The Debye temperature of our compound at 0 GPa and 300~K temperature equals 461.25~K.

\section{Conclusion}

Using first-principles calculations, the half-metallicity has been predicted in $\text{KBeO}_{3}$, with the total magnetic moment of 3$\mu_{\text{B}}$ per unit cell. The magnetism of this material mainly originates from the $p$ orbital of O atoms. The results indicate that $\text{KBeO}_{3}$ is an anisotropic ductile material and keeps a perfect half-metallicity in the pressure range from 0 to 97 GPa. On the other hand, a transition occurs from ferromagnetic to non-magnetic phase under this pressure value. Therefore, $\text{KBeO}_{3}$ is a promising material as a source material for spin injection and electrode materials for Giant Magneto-Resistance (GMR) devices. In addition, the thermodynamic properties are assessed using the quasi-harmonic Debye model.

\clearpage
\ukrainianpart

\title
{Індукований високим тиском магнітний фазовий перехід у напівметалічному перовскіті $\textbf{KBeO}_\textbf{3}$  %
}
\author{M. Хамлат\refaddr{label1}, K. Амара\refaddr{label1}, K. Будіа\refaddr{label2},
        Ф. Хелфаї\refaddr{label1}\, Г. Буталеб\refaddr{label1} }
\addresses{
\addr{label1} Фізико-хімічна лабораторія,  Університет ім. $\textrm{D}^\textrm{r}$ Тахара Мулая м. Саїда, 20000 Саїда, Алжир
\addr{label2} Фізико-хімічна лабораторія новітніх матералів, Університет  Джіллалі Ліабеса, 22000  Сіді-Бель-Аббес, Алжир
}

\makeukrtitle

\begin{abstract}
\tolerance=3000%
У статті досліджуються структурні, механічні, магнітно-електронні і термодинамічні властивості перовскіту  KBeO$_3$. Обчислення здійснено повно-потенціальним  методом приєднаних плоских хвиль, імплементованим в  WIEN2k коді, який базується на теорії функціоналу густини з використанням узагальненого градієнтного наближення. Обчислена енергія формування і пружні константи вказують на здатність синтезуватися і механічну стійкість  KBeO$_3$. Більше того, наші результати показали, що KBeO$_3$ є напівметалічним матеріалом з напівметалічною забороненою зоною   0.67 eV та сумарним магнітним моментом  3$\mu_{\text{B}}$ на елементарну комірку. Окрім цього,  KBeO$_3$ утримує напівметалічний характер під дією тиску аж до  97 GPa, що відповідає передбачуваному переходу магнітної фази з феромагнітного в немагнітний стан. Квазігармонічну модель Дебая використано для аналізу об'ємного коефіцієнта  $V/V_{0}$, об'ємного модуля, питомої теплоємності, температурного розширення та температури Дебая.
\keywords перовскіт, напівметалічний характер, механічна стійкість, перехід магнітної фази, термодинамічні властивості
\end{abstract}

\lastpage
\end{document}